\documentclass[aps,twocolumn,prl,superscriptaddress]{revtex4-1}
\usepackage{amsmath}
\usepackage{amssymb}
\usepackage[final]{graphicx}
\usepackage{dcolumn}
\usepackage{bm}
\usepackage{color}
\usepackage{setspace}

\def \equi#1{\mathrel{\mathop{\kern 0pt\sim}\limits_{#1}}} 
\def \equil#1{\mathrel{\mathop{\kern 0pt\to}\limits_{#1}}} 

\begin{document}

\title{Spontaneous formation of chaotic protrusions in a polymerizing active gel layer}
\author{N. Levernier}
\affiliation{Department of Biochemistry, University of Geneva, Geneva, Switzerland}
\affiliation{Department of Theoretical Physics, University of Geneva, Geneva, Switzerland}

\author{K. Kruse}
\affiliation{Department of Biochemistry, University of Geneva, Geneva, Switzerland}
\affiliation{Department of Theoretical Physics, University of Geneva, Geneva, Switzerland}
\affiliation{National Center of Competence in Research Chemical Biology, University of Geneva,Geneva, Switzerland}

\begin{abstract}
The actin cortex is a thin layer of actin filaments and myosin motors beneath the outer membrane of animal cells. It determines the cells' mechanical properties and forms important morphological structures. Physical descriptions of the cortex as a contractile active gel suggest that these structures can result from dynamic instabilities. However, in these analyses the cortex is described as a two-dimensional layer. Here, we show that the dynamics of the cortex is qualitatively different when gel fluxes in the direction perpendicular to the membrane are taken into account. In particular, an isotropic cortex is then stable for arbitrarily large active stresses. If lateral contractility exceeds vertical contractility, the system can either from protrusions with an apparently chaotic dynamics or a periodic static pattern of protrusions. 
\end{abstract}

 \maketitle

The cytoskeleton is a dynamic meshwork of filamentous protein assemblies that is of crucial importance for living cells~\cite{Bray01,Howard2001,Alberts2008}. The filaments, formed by the proteins actin or tubulin, interact with molecular motors like myosins or kinesins. By using a chemical fuel, these are able to generate mechanical stresses in the filament network  {making it} 
an active gel~\cite{Kruse:2004il}. Physical studies of active matter often use a hydrodynamic approach~\cite{Marchetti:2013bp}. For the cytoskeleton, this approach has been used to study structures and dynamics that are either alien to conventional matter or behave differently from their passive counterparts. Examples include topological point defects~\cite{Kruse:2004il,Sanchez:2012gt,Giomi:2013ky,Thampi:2013cua,Pismen:2013ie,Keber:2014fh,Pearce:2018wt}, spontaneous flow transitions~\cite{Voituriez:2007jy,Furthauer:2012iu,Duclos:2018it}, or buckling instabilities of contracting gels~\cite{Hannezo:2014gp,Ideses:2018dn}.

Physical tools and concepts have also been used to study structures formed by the cytoskeleton in living cells. In this context, the actin cortex -- a thin actomyosin layer beneath the plasma membrane of animal cells -- is of particular interest. It determines the mechanical properties of animal cells~\cite{Fritzsche:2016fx,Chugh:2017de} and plays an important role in cellular morphogenesis~\cite{Salbreux:2012bo}. In addition, the cortex hosts a number of 
cytoskeletal structures~\cite{Blanchoin:2014jr}. A particularly {striking} 
example is the contractile actomyosin ring that cleaves animal cells into two daughter cells during 
division. The ring might be generated by a dynamic instability of the cortex~\cite{Zumdieck:2005wba,Salbreux:2009fp} and its dynamics when connected to a membrane have been studied~\cite{Sedzinski:2011ef,Turlier:2014hq}.

The cortex is formed by actin filaments that polymerize directly at the cell membrane~\cite{Alberts2008}. Its thickness, which has been reported to be of a few hundred nanometers~\cite{Clark:2013ef,Clausen:2017jc}, is determined by the actin polymerization velocity and its disassembly rate~\cite{Joanny:2013fm}. Due to the large aspect ratio, in theoretical analysis, the cortex is typically considered as a two-dimensional surface~\cite{Mayer:2010kt,Naganathan:2014fc,Berthoumieux:2014eo,Mietke:2019ki}. However, a formal reduction of the dynamics to the two lateral dimensions as in a thin-film approximation is not always possible.   

In this work, we explore the dynamics of a layer of a contractile active gel on a rigid substrate with a constant influx of material {at the substrate.} 
Contrary to previous work, we also account for gel flows in the dimension perpendicular to the substrate. We find that an isotropic layer is always stable against perturbations. In case active contractile stresses are stronger in the lateral direction than perpendicular to the substrate, we observe an instability leading to a local increase in gel density. {This is} similar to the dynamics {when neglecting the gel thickness}
~\cite{Bois:2011kx,Mietke:2019ki}. However, the states emerging in our system differ fundamentally from those reported {in these works}: we find periodic stationary states and states that are apparently chaotic. In {all} 
cases, they exhibit a characteristic length scale. 
{When neglecting the gel thickness, such states can be obtained only after modifying the dynamics, for example, by introducing non-linear gel assembly terms~\cite{Hannezo:2015ba}.} 

\begin{figure}
\includegraphics[width=0.45\textwidth]{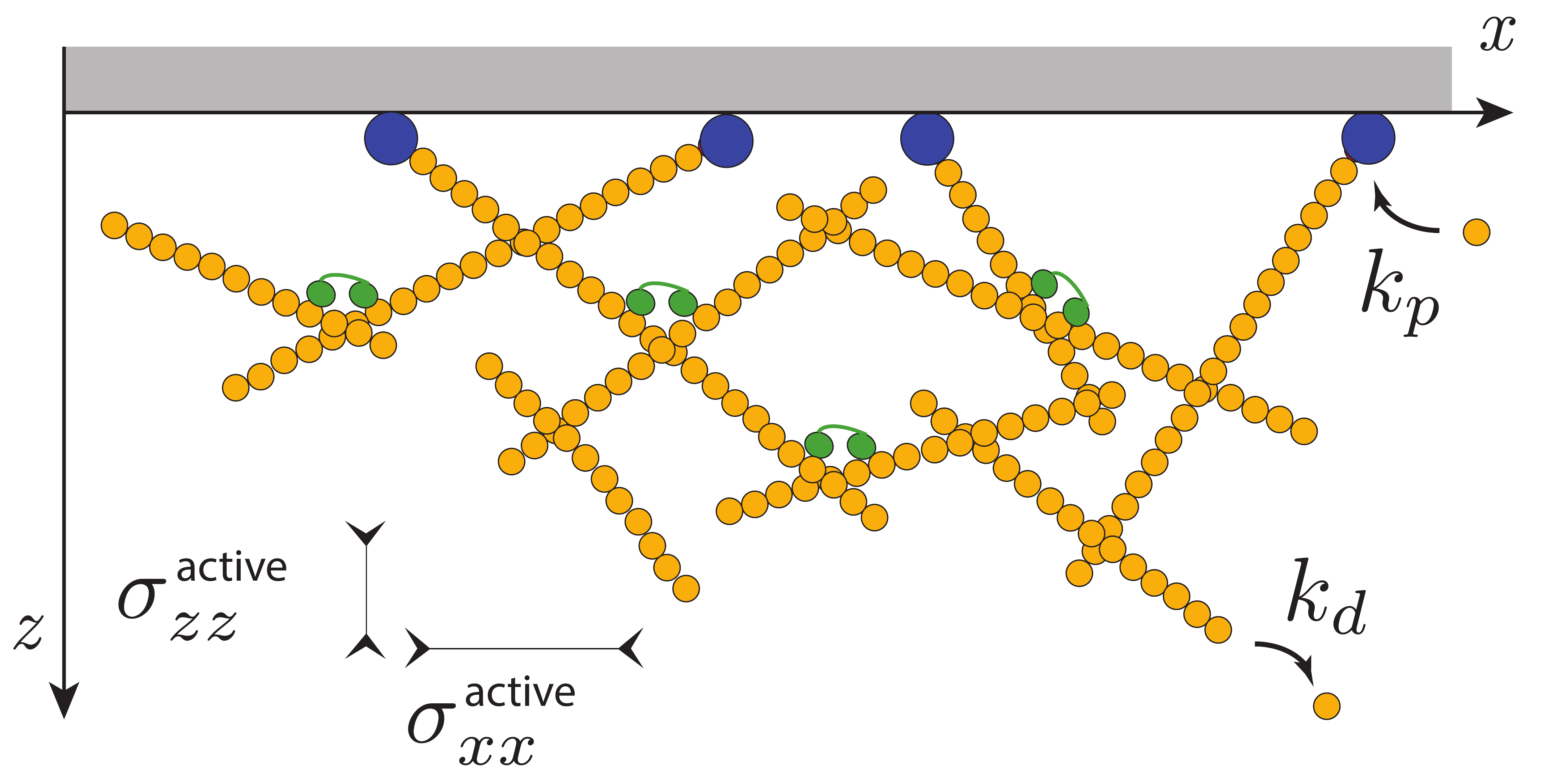}
\caption{{(color online)} Illustration of an anisotropic active gel polymerizing on a surface. Assembly of the filaments ({small yellow circles}) at $z=0$ is mediated by nucleators ({big blue circles}) of density $\rho_0$. This leads to a flux of gel $k_p \delta\rho_0$ perpendicular to the surface, where $k_p$ is the polymerization rate and $\delta$ the length added to a filament upon addition of a subunit. The gel disassembles in the bulk at a constant rate $k_d$. Active stress is generated by motor proteins ({green connected circles}).}
\label{sketch}
\end{figure}
We consider a three-dimensional viscous active gel, growing into the half space $z\ge0$ at the surface $(x,y,z=0)$, Fig.~\ref{sketch}. For the sake of simplicity, we assume invariance along the direction $y$. The state of the gel is determined by the density $\rho$ and {the velocity field} 
$\mathbf{v}=(v_x,v_z)$. Mass conservation and force balance yield:
\begin{align}
\partial_t \rho + \partial_x (\rho v_x)+ \partial_z (\rho v_z)&=-k_d \rho \label{eq:massConservation}\\
\eta \,[2 \partial_{xx} v_x + \partial_{xz} v_z + \partial_{zz} v_x]&= \partial_x \Pi_x(\rho) \label{eq:forceBalanceX} \\
\eta \,[2 \partial_{zz} v_z + \partial_{xz} v_x + \partial_{xx} v_z]&= \partial_z \Pi_z(\rho)  \label{eq:forceBalanceZ}.
\end{align}
In these equations, $k_d$ is the gel's disassembly rate, $\eta$ denotes the viscosity and $\Pi_{x,z}$ are the components of the non-viscous contribution to the total stress, which we assume to have only diagonal components. {It} 
has two contributions: an effective hydrostatic pressure and the stress generated by active processes in the gel. 
In general, both contributions could be anisotropic due to local filament alignment. For simplicity, we consider an isotropic hydrostatic pressure and an active stress with fixed anisotropy. We discuss the limitations of this assumption at the end of our manuscript. Similar to Ref.~\cite{Joanny:2013fm}, we choose
\begin{align}
\label{eq:effectivePressure}
\Pi_{x,z}(\rho)=-a_{x,z}\rho^3 + b \rho^4
\end{align}
with $b>0$. We take $a_{x,z}>0$ reflecting the contractile nature of the active stress. Consequently, $\Pi_x$ and $\Pi_z$ decrease with $\rho$  for low enough densities and increase with $\rho$ as $\rho\to\infty$. We checked that our results do not depend on the specific form chosen for $\Pi_{x,z}$. It suffices that the dependence of $\Pi_{x,z}$ on $\rho$ {has a shape similar to that given in Eq.~(\ref{eq:effectivePressure})}.

The dynamic equations are completed with the boundary condition $\rho(z=0)=\rho_0$ and $v_z(z=0)=v_p$, where $\rho_0$ is the density of nucleators at the boundary and $v_p$ the polymerization speed. In addition, the tangential stress at the surface is balanced by friction of the gel along the membrane:
\begin{align}
\left.\eta \, \partial_z v_x\right|_{z=0}  &= \xi v_x(z=0),
\end{align}
where $\xi$ is the friction constant. Furthermore, we impose periodic boundary conditions in the $x$-direction with period $L$. In the $z$-direction, the system extends to infinity and $\rho\to0$ for $z\to\infty$.  

If we ignore the $z$-direction in Eqs.~\eqref{eq:massConservation}-\eqref{eq:forceBalanceZ}, we arrive at a dynamic system similar to others that have been used before in the analysis of the actin cortex or other thin active gel layers~\cite{Bois:2011kx,Mietke:2019ki}. For high enough activity, these systems typically generate contracted stationary states with a single region of high gel density {unless non-linear assembly terms are considered~\cite{Hannezo:2015ba}}. {However, there is no way to integrate the dynamic equations \eqref{eq:massConservation}-\eqref{eq:forceBalanceZ} over $z$ to get a closed system for the averaged fields $\bar{\rho}(x)$ and $\bar{v}(x)$ in one dimension: a formal integration over $z$ creates term such as $\int v_x \rho\,\mathrm{d}z$, $\int \Pi(\rho)\,\mathrm{d}z$ and boundary terms such as $\partial_z v_x |_{z=0}$ or $\partial_x v_x |_{z=0}$. These terms cannot be readily expressed  in terms of $\bar{\rho}(x)$ and $\bar{v}(x)$ for a compressible system}.

In the following we will use a dimensionless form of the dynamic equations~\eqref{eq:massConservation}-\eqref{eq:effectivePressure} and scale time by $k_d^{-1}$, length by $\ell=v_p k_d^{-1}$,  concentration by $\rho_0$, and stress by $\eta k_d$. Hence, the relevant parameters are $a_{x,z} (k_d\eta)^{-1}\rho_0^{3}$, $b (k_d \eta)^{-1}\rho_0^{4}$, and $\xi v_p (k_d\eta)^{-1}$. We will use the same notation for the original and the rescaled parameters. For future use, we introduce the anisotropy parameter $\epsilon=a_x/a_z$.

In case the gel is invariant with respect to lateral translations, two qualitatively different stationary states exist: either the density decays exponentially or it jumps from a finite value $\rho_e$ to zero at $z=h\sim \ell$~\cite{Joanny:2013fm}. For the "exponential profile", the velocity $v_z$ converges to a finite value for $z\to\infty$. In contrast, $v_z=0$ for $z>h$ for the "step profile". The latter {exists} 
only 
for large enough contractility, that is, for 
$a_z$ above a critical value $a_c\ge0${, and when} 
the density of nucleators $\rho_0$ 
exceeds a critical value $\rho_c$. Otherwise, the profile decays exponentially. 

We start our analysis of the general case by numerically solving the dynamic equations (\ref{eq:massConservation})-(\ref{eq:effectivePressure}). In each time step, we first determine the velocity field through the force balance equations~(\ref{eq:forceBalanceX}) and (\ref{eq:forceBalanceZ}), where we use Fourier decomposition along $x$ and finite-differences along $z$. We then update the density. 
The contribution of $\partial_z(\rho v_z)$ is obtained by an up-wind finite-differences scheme in real space. To improve the stability of the scheme, we have added a diffusion term with diffusion constant $D=10^{-3}$ to the mass conservation equation~(\ref{eq:massConservation}). In all simulations, we use $\Delta x=0.004$, $\Delta z=0.007$, and $\Delta t=0.0005$. We checked that our results do not change for smaller values. As initial condition, we take $\rho(x_i,z_j) = \rho_0(z_j)(1+\tilde\eta(x_i,z_j))$, where $\tilde\eta(x_i,z_j)$ are independent and uniformly distributed random numbers between $-0.05$ and $0.05$.

\begin{figure}
\includegraphics[width=0.45\textwidth]{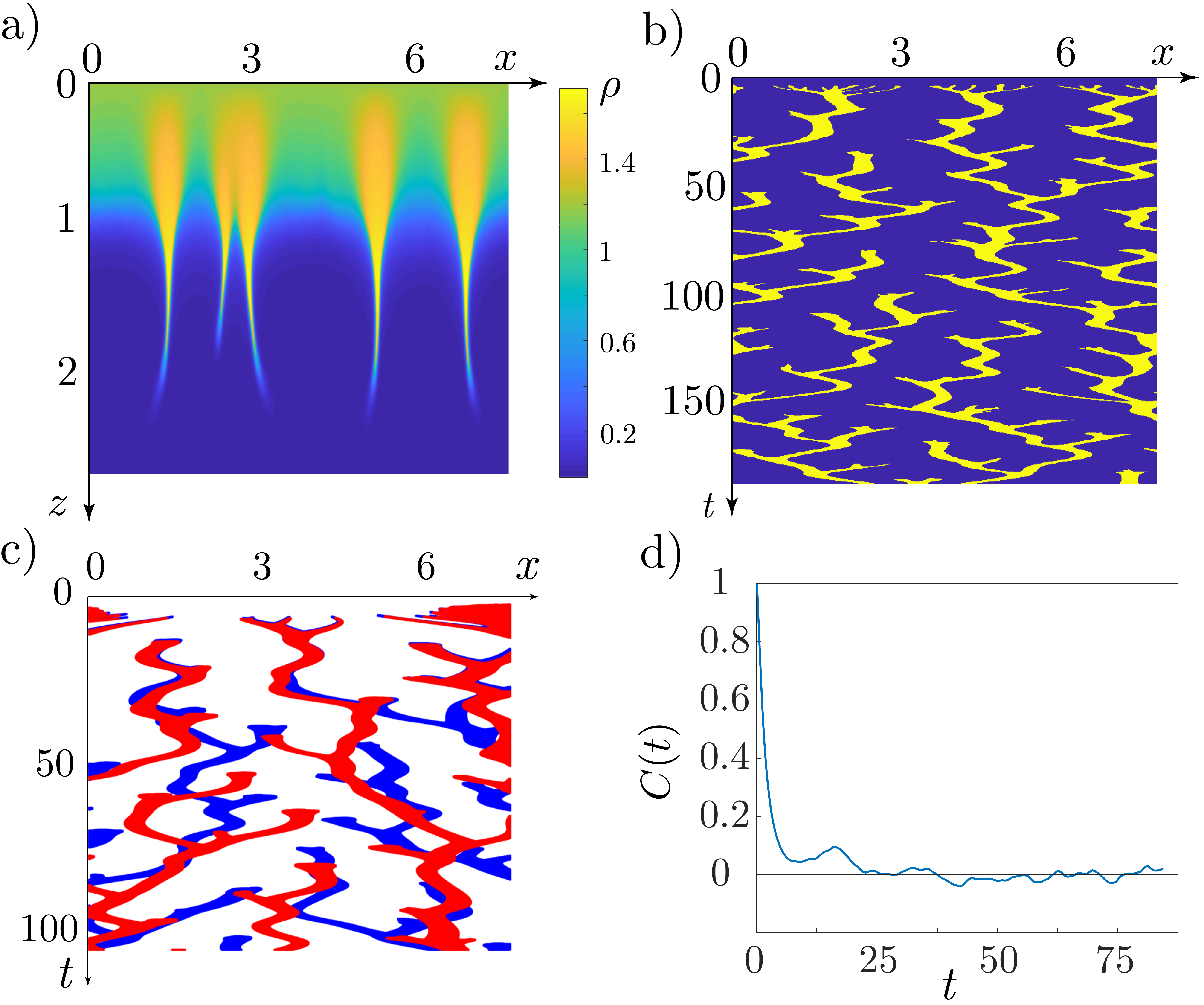}
\caption{{(color online)} Chaotic state of continuously appearing and merging protrusions in the case of low friction. a) Snapshot of the density field $\rho$. b) Kymograph of the density at $z=h$ with values $\rho>\rho_e$ shown in yellow {(bright)}. Parameters are $L=8.2$, $\epsilon=1.25$, $a_z=7$, $b=4.9$, $\xi v_p/(k_d\eta)=0$ and $\rho_0=\rho_e=1.18$. {c) Two kymographs with close initial conditions, that depart one from the other as a signature of chaos (same parameters as in (b)). d) Normalized protrusion auto-correlation function for the solution shown in (b), $C(t)=\langle (\sigma(x,\tau)-\langle \sigma\rangle)(\sigma(x,\tau+t)-\langle \sigma\rangle) \rangle$, where $\sigma(x,t)$ is 1 inside a protrusion and 0 outside and averages are performed over time and space.   } }
\label{fig:chaotic}
\end{figure}

We first consider the limit of zero friction, i.e., $ \partial_z v_x(z=0)=0$. If $a_z>a_c$, then there is a critical anisotropy $\epsilon_c$, such that the step profile is unstable for $\epsilon>\epsilon_c$. In Figure~\ref{fig:chaotic} and Movie~1~\cite{SM}, we present the solution for $a_z=7>a_c$ and $\epsilon=1.18>\epsilon_c$. Locally, the gel contracts laterally, which leads to the formation of finger-like protrusions growing into the direction of increasing $z$. 
{Continuously,} neighboring protrusions merge and new ones appear. 
As can be seen in the kymograph, Fig.~\ref{fig:chaotic}, the system does not settle in a periodic state. 
Instead the dynamics appears to be chaotic. {The chaotic nature of our system is highlighted by its sensitivity to initial conditions, Fig.~\ref{fig:chaotic}c, and the relaxation of the protrusion auto-correlation $C(t)$, Fig.~\ref{fig:chaotic}d, with 
$C(t)=\langle (\sigma(x,\tau)-\langle \sigma\rangle)(\sigma(x,\tau+t)-\langle \sigma\rangle) \rangle$, where $\sigma(x, t)$ is 1 inside a protrusion and 0 outside. Furthermore,  $\langle\ldots\rangle$ denotes a space and time average. } Our numerical results indicate that the chaotic state presented in Fig.~\ref{fig:chaotic} emerges directly at the critical anisotropy $\epsilon_c$. Such a direct transition to chaotic dynamics, which is impossible for systems with a finite number of degrees of freedom, is known to exist for other spatially extended dynamic systems~\cite{Aranson:2002vf,Tribelsky:1996do}. 

\begin{figure}
\includegraphics[width=0.45\textwidth]{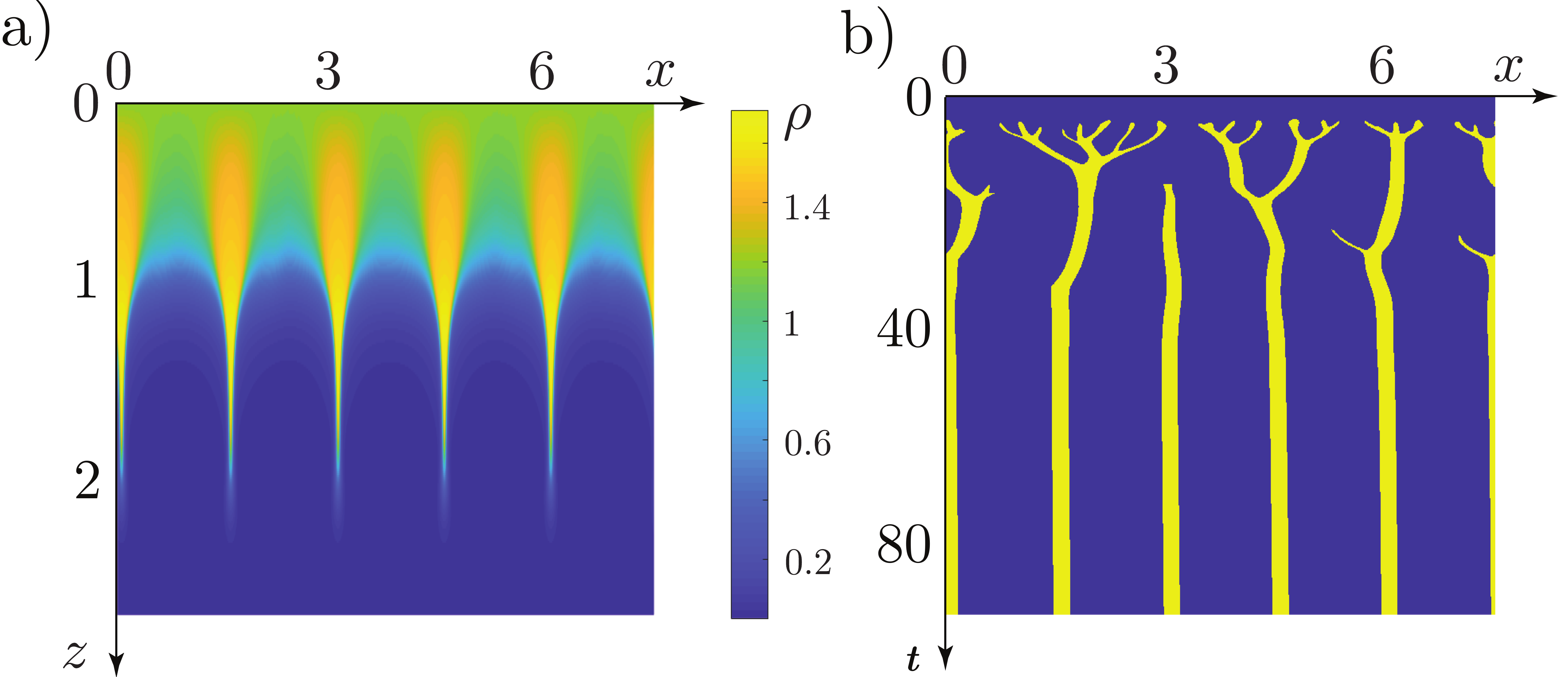}
\caption{Stationary periodic pattern in the case of high friction. a) Snapshot of the density field $\rho$. b) Kymograph of the density at $z=h$ with values $\rho>\rho_e$ shown in yellow {(bright)}. Parameters are $L=8.2$, $\epsilon=1.25$, $a_z=7$, $b=4.9$, $\xi v_p/(k_d\eta)=5$ and $\rho_0=\rho_e=1.18$.}
\label{fig:stationaryProtrusion}
\end{figure}

Note that for $\xi=0$, $\ell$ is the only length scale present in our system. 
{It} fixes the order of the density of protrusions {and} 
the thickness of the step profile. When reducing the lateral extension $L$ sufficiently, the system forms a single protrusion for $\epsilon>\epsilon_c$, which is now stationary. In that case, new material that is introduced into the system at $z=0$ is drawn into the protrusion by the contractile lateral activity and cannot form a new protrusion. {Also} in the case of large enough friction $\xi$, a stationary state emerges at the instability. It consists of a periodic arrangement of protrusions, Fig.~\ref{fig:stationaryProtrusion} and Movie~2~\cite{SM}.

{We} perform a linear stability analysis of the step profile with $\rho_0=\rho_e$. Then $\rho(x,z)=\rho_0$ for $0\le z\le h$ and zero otherwise. Note that this state only exists if $a_z\ge a_c$. Let $\alpha=\frac{d\Pi_x}{d\rho}|_{\rho_e}$ and $\beta=\frac{d\Pi_z}{d\rho}|_{\rho_e}$. {Let us recall that for obtaining the step profile, we necessarily have  $\beta>0$~\cite{Joanny:2013fm}.} Then, the growth rate $s$ of a perturbation $\tilde{\rho} \exp (i q_x x + i q_z z)$ for $0\le z\le h$ is given by 
\begin{equation}
s=-\frac{\rho_e}{2} \left(\beta+(\alpha-\beta)\frac{q_x^2}{q_x^2+q_z^2}\right) 
\label{growthrate}
\end{equation}
with real wave numbers $q_x$ and $q_z$. {Note that the expression for $s$ is independent of the value of the friction coefficient $\xi$. Its value only affects the range of values of $q_z$ that satisfy the boundary condition at $z=0$.}

From Equation~(\ref{growthrate}), the step profile is unstable if $\alpha$ is negative. {Together with $\beta>0$ this implies $a_x>a_z$, i.e., an anisotropic active stress that is more contractile in the $x$-direction.} In this case, a local increase of the gel density leads to a local increase of the contractility in $x$-direction. In turn, this generates flows toward the region of increased density, leading to a further increase. The most unstable modes $\mathbf{q}=(q_x,q_z)$ are $(q_x,0)$ for any $q_x$ and $(\infty,q_z)$ for any finite $q_z$. 
{Contrary to the usual case with one or a few dominant unstable wavelengths, where 
a pattern with a corresponding characteristic length scale emerges, here} an infinite number of modes becomes simultaneously unstable. {In this case,} the steady state can lose stability directly in favor of a chaotic state as our numeric solution suggests for the parameter values corresponding to Fig.~\ref{fig:chaotic}. 

\begin{figure}%
\includegraphics[width=0.4\textwidth]{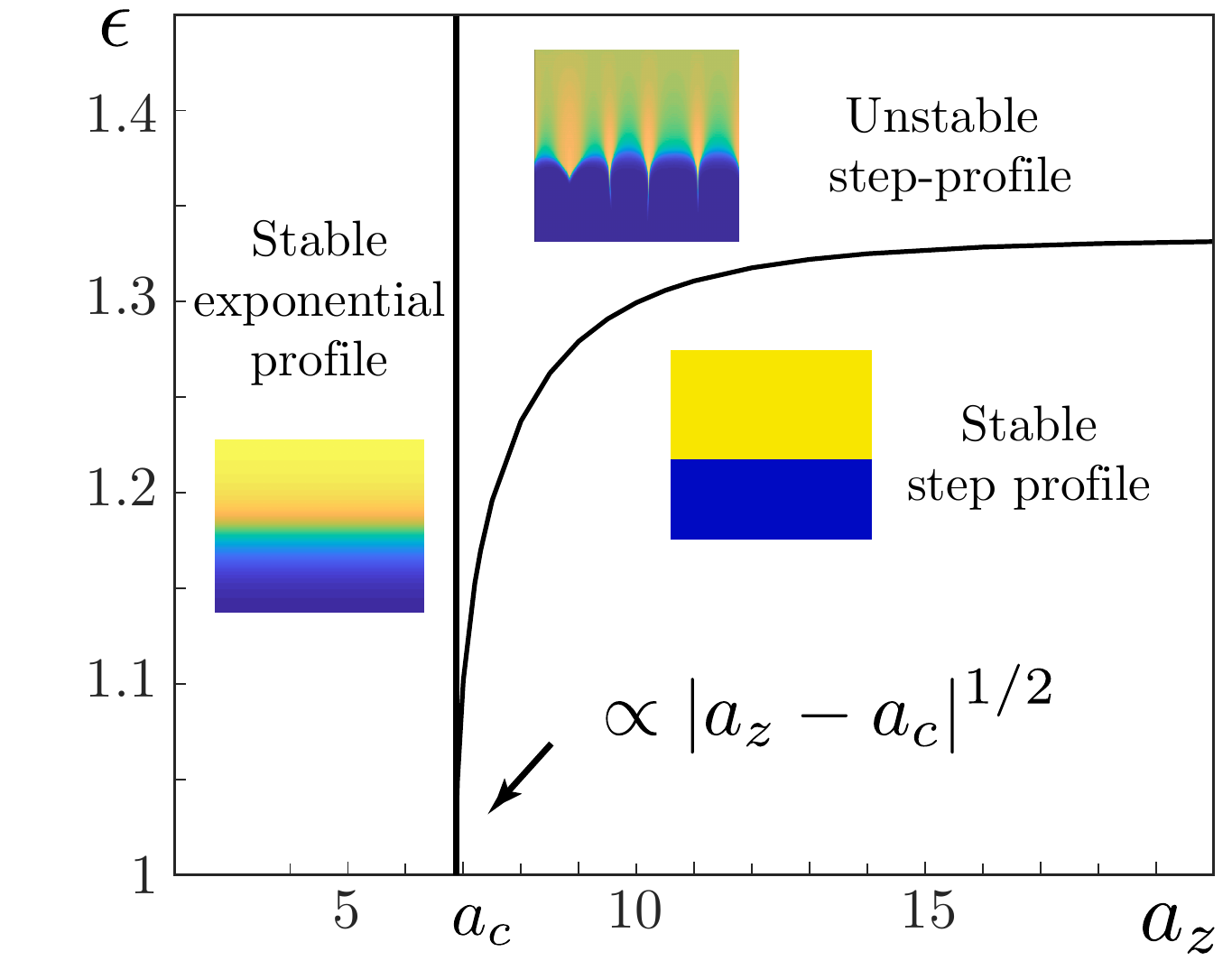}
\caption{Stability diagram. For $a_z<a_c$ with $a_c$ denoting the critical activity for the transition from an exponential decaying density to one jumping to zero at $z=h$~\cite{Joanny:2013fm}, the steady state is linearly stable. For $a_z>a_c$ the steady state is linearly stable as long as the anisotropy $\epsilon$ is below a critical value. Parameters are $b =4.9$ and $\rho_0=\rho_e$ for $a_z>a_c$. For $a_z<a_c$, the system evolves into an exponential profile independently of the value of $\rho_0$. The value of the friction $\xi v_p/(k_d\eta)$ is irrelevant for the stability.}
\label{fig:phasediagram}
\end{figure}

In Figure~\ref{fig:phasediagram}, we present the stability diagram in the plane of the activity parameter $a_z$ and the anisotropy $\epsilon$. Even though the precise form of the stability boundary $\epsilon_c(a_z)$ depends on the detailed expression of the functions $\Pi_{x,z}$, we always have $\epsilon_c-1\propto\sqrt{|a_z - a_c|}$ for $a_z\gtrsim a_c$. Indeed, $\frac{d\Pi_z}{d\rho}|_{\rho_e^c}=0$, where $\rho_e^c\equiv\rho_e$ for $a_z=a_c$, so that for $a_z=a_c+\delta a$ we get $\rho_e-\rho_e^c \propto \sqrt{\delta a}$. Furthermore, $\epsilon_c$ saturates for $a_z\to\infty$ independently of the precise from of $\Pi_{x,z}$. 
{For} $\Pi_{x,z}$ 
given by Eq.~(\ref{eq:effectivePressure}), we find $\epsilon=4b\rho_e/a_z=O(1)$ for $a_z\to \infty$. 
 
The exponential profiles are unconditionally stable. The crucial difference compared to the step profiles is that, here, the steady-state velocity $v_z(z)$ reaches a non-zero velocity $v_\infty$ when $z \to \infty$. Hence, any perturbation $\delta \rho$ will be transported at a finite velocity away from the substrate at $z=0$. As the gel degrades with time at rate $k_d$, the perturbation will reach a region where the density $\rho$ is exponentially small in finite time. The perturbation then satisfies the advection-degradation equation $\partial_t \delta\rho + v_\infty \partial_z \delta\rho =-k_d \delta\rho$, which implies that it eventually disappears.

{Since there is not a single dominant 
mode that grows fastest, the linear stability analysis 
cannot yield information about the state after the instability. Even to study whether it is stationary or not requires a non-linear analysis. This is challenging as the standard techniques of weakly non-linear analysis cannot be applied in our case. Instead, we can give a hand-waving reasoning for distinguishing between stationary and chaotic states. Let us consider a stationary periodic pattern $\rho^\mathrm{per}(x,z)$ to which we add a perturbation  $\delta \rho$. By expanding the mass-balance equation close to the surface, we see that the fate of $\delta\rho$ is determined by the competition of two transport processes. On one hand, potentially unstable terms arise due to $v_x^\mathrm{per}$ and $\delta v_x$, describing the attraction of neighboring protrusions thanks to contractile effects. On the other hand, stabilizing terms emerge due to the transport of material towards increasing $z$ at velocity $v_p$. Even though $v_x^\mathrm{per}$ and $\delta v_x$ cannot be explicitly determined, the boundary condition implies that they both are of order $v_\mathrm{approach}\equiv v_p\eta(\ell\xi)^{-1}$. The transition between the chaotic and the periodic pattern occurs when $v_p$ and $v_\mathrm{approach}$ are of the same order, that is, for $\eta/\xi\sim\ell$. Indeed,  $0.2=v_\mathrm{approach}<v_p$ in Fig.~\ref{fig:stationaryProtrusion}, whereas $\infty=v_\mathrm{approach}>v_p$ in  Fig.~\ref{fig:chaotic}.}


In this letter, we have shown that the dynamics of an active gel polymerizing at a surface is fundamentally different from the dynamics of this system, when fluxes perpendicular to the surface are neglected. Hence, results obtained within the thin-layer approximation have to be taken with caution. 

Let us comment on our assumption of fixed anisotropy of the active gel. In the cytoskeleton, the differences in the active stress in the direction lateral and perpendicular to the surface result from the alignment of the filaments along the membrane. In a full description of the cytoskeleton, the filament alignment is a dynamic field and the anisotropy should emerge spontaneously. Filament alignment {induced by contraction} 
is a natural feature of the dynamics of active polar or nematic gels~\cite{Salbreux:2009fp}. In addition, the presence of a surface should also lead to a preferential alignment of the filaments at the boundary. We do thus not expect the dynamics to change qualitatively in presence of a dynamic orientational order field and postpone its analysis to future work.

Compared to previous analyses of thin layers of active gels, we found 
a direct transition to a chaotic state and a transition controlled by friction. Periodic patterns also emerge in multi-component active systems~\cite{Kruse:2003vr,Bois:2011kx,Gowrishankar:2012ek,Kumar:2014fx,Banerjee:2017ja}. {However, there the period results from balancing friction and diffusion, which is different from our system, where it is a consequence of filament turnover.} Spatiotemporal chaos in absence of inertial effects has been reported in active systems~\cite{Dombrowski:2004eu,Sanchez:2012gt,Neef:2014ez,Ramaswamy:2016kb,Doostmohammadi:2016bd}. However, these system were incompressible, did not exhibit turnover, and displayed dynamic nematic or polar order. In the two-component system studied in Ref.~\cite{Doostmohammadi:2016bd} {or in presence of non-linear gel assembly terms~\cite{Hannezo:2015ba}}, a transition between a static periodic and a chaotic state was obtained by changing the friction coefficient. However, in these cases the friction permeated the whole system, whereas in our case, it is restricted to a boundary. Together, these features show that adding a dimension can have effects similar to the introduction of additional physical properties.

In addition to the activity and the friction, we can also use $\rho_0$ as a control parameter. In a living cell, this parameter is set by the density of active actin-filament nucleators. In this way the cell has a readily accessible controller to switch between different morphologies. {Although a direct observation of this mechanism inside living cells may be tricky, 
some known phenomena could be 
manifestations of this instability. First, a ``ramified actin network'' consisting of actin bundles emerging from the cortex has been observed in spreading Jurkat T cells~\cite{Fritzsche:2017ji}. Second, on the surface of suspended Jurkat T cells, short finger-like protrusions emerge continuously~\cite{Fritzsche:2017ji}. Finally, many cells present ``blebs'' at their surface, which are membrane bulges in regions with a weakened cortex~\cite{Charras:2008uu}. In our system, the cortical density is reduced in the regions between the protrusions, which offers a possible mechanism for the appearance of blebs.} In upcoming work, we will analyse the dynamics of the actin cortex in case the {substrate} 
is deformable 
{to investigate the connection between the cellular structures just mentioned and the cortical instabilities reported in this work}.  

\acknowledgments We acknowledge financial support by the Swiss National Science Foundation through grant number 205321\_175996. N.L. acknowledges support through a Fondation Bettencourt-Schueller prize.


\begin{thebibliography}{42}%
\makeatletter
\providecommand \@ifxundefined [1]{%
 \@ifx{#1\undefined}
}%
\providecommand \@ifnum [1]{%
 \ifnum #1\expandafter \@firstoftwo
 \else \expandafter \@secondoftwo
 \fi
}%
\providecommand \@ifx [1]{%
 \ifx #1\expandafter \@firstoftwo
 \else \expandafter \@secondoftwo
 \fi
}%
\providecommand \natexlab [1]{#1}%
\providecommand \enquote  [1]{``#1''}%
\providecommand \bibnamefont  [1]{#1}%
\providecommand \bibfnamefont [1]{#1}%
\providecommand \citenamefont [1]{#1}%
\providecommand \href@noop [0]{\@secondoftwo}%
\providecommand \href [0]{\begingroup \@sanitize@url \@href}%
\providecommand \@href[1]{\@@startlink{#1}\@@href}%
\providecommand \@@href[1]{\endgroup#1\@@endlink}%
\providecommand \@sanitize@url [0]{\catcode `\\12\catcode `\$12\catcode
  `\&12\catcode `\#12\catcode `\^12\catcode `\_12\catcode `\%12\relax}%
\providecommand \@@startlink[1]{}%
\providecommand \@@endlink[0]{}%
\providecommand \url  [0]{\begingroup\@sanitize@url \@url }%
\providecommand \@url [1]{\endgroup\@href {#1}{\urlprefix }}%
\providecommand \urlprefix  [0]{URL }%
\providecommand \Eprint [0]{\href }%
\providecommand \doibase [0]{http://dx.doi.org/}%
\providecommand \selectlanguage [0]{\@gobble}%
\providecommand \bibinfo  [0]{\@secondoftwo}%
\providecommand \bibfield  [0]{\@secondoftwo}%
\providecommand \translation [1]{[#1]}%
\providecommand \BibitemOpen [0]{}%
\providecommand \bibitemStop [0]{}%
\providecommand \bibitemNoStop [0]{.\EOS\space}%
\providecommand \EOS [0]{\spacefactor3000\relax}%
\providecommand \BibitemShut  [1]{\csname bibitem#1\endcsname}%
\let\auto@bib@innerbib\@empty
\bibitem [{\citenamefont {Bray}(2001)}]{Bray01}%
  \BibitemOpen
  \bibfield  {author} {\bibinfo {author} {\bibfnamefont {D.}~\bibnamefont
  {Bray}},\ }\href@noop {} {\emph {\bibinfo {title} {{Cell Movements}}}},\
  \bibinfo {edition} {2nd}\ ed.,\ From Molecules to Motility\ (\bibinfo
  {publisher} {Taylor {\&} Francis},\ \bibinfo {year} {2001})\BibitemShut
  {NoStop}%
\bibitem [{\citenamefont {Howard}(2001)}]{Howard2001}%
  \BibitemOpen
  \bibfield  {author} {\bibinfo {author} {\bibfnamefont {J.}~\bibnamefont
  {Howard}},\ }\href@noop {} {\emph {\bibinfo {title} {{Mechanics of Motor
  Proteins and the Cytoskeleton}}}}\ (\bibinfo  {publisher} {Sinauer
  Associates},\ \bibinfo {address} {Sunderland},\ \bibinfo {year}
  {2001})\BibitemShut {NoStop}%
\bibitem [{\citenamefont {Alberts}\ \emph {et~al.}(2008)\citenamefont
  {Alberts}, \citenamefont {Johnson}, \citenamefont {Lewis}, \citenamefont
  {Raff}, \citenamefont {Roberts},\ and\ \citenamefont {Walter}}]{Alberts2008}%
  \BibitemOpen
  \bibfield  {author} {\bibinfo {author} {\bibfnamefont {B.}~\bibnamefont
  {Alberts}}, \bibinfo {author} {\bibfnamefont {A.}~\bibnamefont {Johnson}},
  \bibinfo {author} {\bibfnamefont {J.}~\bibnamefont {Lewis}}, \bibinfo
  {author} {\bibfnamefont {M.}~\bibnamefont {Raff}}, \bibinfo {author}
  {\bibfnamefont {K.}~\bibnamefont {Roberts}}, \ and\ \bibinfo {author}
  {\bibfnamefont {P.}~\bibnamefont {Walter}},\ }\href@noop {} {\emph {\bibinfo
  {title} {{Molecular Biology of The Cell}}}},\ \bibinfo {edition} {5th}\ ed.,\
  edited by\ \bibinfo {editor} {\bibfnamefont {B.}~\bibnamefont {Alberts}}\
  (\bibinfo  {publisher} {Garland Science},\ \bibinfo {year}
  {2008})\BibitemShut {NoStop}%
\bibitem [{\citenamefont {Kruse}\ \emph {et~al.}(2004)\citenamefont {Kruse},
  \citenamefont {Joanny}, \citenamefont {J{\"u}licher}, \citenamefont {Prost},\
  and\ \citenamefont {Sekimoto}}]{Kruse:2004il}%
  \BibitemOpen
  \bibfield  {author} {\bibinfo {author} {\bibfnamefont {K.}~\bibnamefont
  {Kruse}}, \bibinfo {author} {\bibfnamefont {J.-F.}\ \bibnamefont {Joanny}},
  \bibinfo {author} {\bibfnamefont {F.}~\bibnamefont {J{\"u}licher}}, \bibinfo
  {author} {\bibfnamefont {J.}~\bibnamefont {Prost}}, \ and\ \bibinfo {author}
  {\bibfnamefont {K.}~\bibnamefont {Sekimoto}},\ }\href@noop {} {\bibfield
  {journal} {\bibinfo  {journal} {Phys. Rev. Lett.}\ }\textbf {\bibinfo
  {volume} {92}},\ \bibinfo {pages} {078101} (\bibinfo {year}
  {2004})}\BibitemShut {NoStop}%
\bibitem [{\citenamefont {Marchetti}\ \emph {et~al.}(2013)\citenamefont
  {Marchetti}, \citenamefont {Joanny}, \citenamefont {Ramaswamy}, \citenamefont
  {Liverpool}, \citenamefont {Prost}, \citenamefont {Rao},\ and\ \citenamefont
  {Simha}}]{Marchetti:2013bp}%
  \BibitemOpen
  \bibfield  {author} {\bibinfo {author} {\bibfnamefont {M.~C.}\ \bibnamefont
  {Marchetti}}, \bibinfo {author} {\bibfnamefont {J.~F.}\ \bibnamefont
  {Joanny}}, \bibinfo {author} {\bibfnamefont {S.}~\bibnamefont {Ramaswamy}},
  \bibinfo {author} {\bibfnamefont {T.~B.}\ \bibnamefont {Liverpool}}, \bibinfo
  {author} {\bibfnamefont {J.}~\bibnamefont {Prost}}, \bibinfo {author}
  {\bibfnamefont {M.}~\bibnamefont {Rao}}, \ and\ \bibinfo {author}
  {\bibfnamefont {R.~A.}\ \bibnamefont {Simha}},\ }\href@noop {}{\bibfield
  {journal} {\bibinfo  {journal} {Rev. Mod. Phys.}\ } \textbf
  {\bibinfo {volume} {85}},\ \bibinfo {pages} {1143} (\bibinfo {year}
  {2013})}\BibitemShut {NoStop}%
\bibitem [{\citenamefont {Sanchez}\ \emph {et~al.}(2012)\citenamefont
  {Sanchez}, \citenamefont {Chen}, \citenamefont {DeCamp}, \citenamefont
  {Heymann},\ and\ \citenamefont {Dogic}}]{Sanchez:2012gt}%
  \BibitemOpen
  \bibfield  {author} {\bibinfo {author} {\bibfnamefont {T.}~\bibnamefont
  {Sanchez}}, \bibinfo {author} {\bibfnamefont {D.~T.~N.}\ \bibnamefont
  {Chen}}, \bibinfo {author} {\bibfnamefont {S.~J.}\ \bibnamefont {DeCamp}},
  \bibinfo {author} {\bibfnamefont {M.}~\bibnamefont {Heymann}}, \ and\
  \bibinfo {author} {\bibfnamefont {Z.}~\bibnamefont {Dogic}},\ }\href@noop {}
  {\bibfield  {journal} {\bibinfo  {journal} {Nature}\ }\textbf {\bibinfo
  {volume} {491}},\ \bibinfo {pages} {431} (\bibinfo {year}
  {2012})}\BibitemShut {NoStop}%
\bibitem [{\citenamefont {Giomi}\ \emph {et~al.}(2013)\citenamefont {Giomi},
  \citenamefont {Bowick}, \citenamefont {Ma},\ and\ \citenamefont
  {Marchetti}}]{Giomi:2013ky}%
  \BibitemOpen
  \bibfield  {author} {\bibinfo {author} {\bibfnamefont {L.}~\bibnamefont
  {Giomi}}, \bibinfo {author} {\bibfnamefont {M.~J.}\ \bibnamefont {Bowick}},
  \bibinfo {author} {\bibfnamefont {X.}~\bibnamefont {Ma}}, \ and\ \bibinfo
  {author} {\bibfnamefont {M.~C.}\ \bibnamefont {Marchetti}},\ }\href@noop {}
  {\bibfield  {journal} {\bibinfo  {journal} {Phys. Rev. Lett.}\ }\textbf
  {\bibinfo {volume} {110}},\ \bibinfo {pages} {228101} (\bibinfo {year}
  {2013})}\BibitemShut {NoStop}%
\bibitem [{\citenamefont {Thampi}\ \emph {et~al.}(2013)\citenamefont {Thampi},
  \citenamefont {Golestanian},\ and\ \citenamefont {Yeomans}}]{Thampi:2013cua}%
  \BibitemOpen
  \bibfield  {author} {\bibinfo {author} {\bibfnamefont {S.~P.}\ \bibnamefont
  {Thampi}}, \bibinfo {author} {\bibfnamefont {R.}~\bibnamefont {Golestanian}},
  \ and\ \bibinfo {author} {\bibfnamefont {J.~M.}\ \bibnamefont {Yeomans}},\
  }\href@noop {} {\bibfield  {journal} {\bibinfo  {journal} {Phys. Rev. Lett.}\
  }\textbf {\bibinfo {volume} {111}},\ \bibinfo {pages} {118101} (\bibinfo
  {year} {2013})}\BibitemShut {NoStop}%
\bibitem [{\citenamefont {Pismen}(2013)}]{Pismen:2013ie}%
  \BibitemOpen
  \bibfield  {author} {\bibinfo {author} {\bibfnamefont {L.~M.}\ \bibnamefont
  {Pismen}},\ }\href@noop {} {\bibfield  {journal} {\bibinfo  {journal} {Phys.
  Rev. E}\ }\textbf {\bibinfo {volume} {88}},\ \bibinfo {pages} {050502(R)}
  (\bibinfo {year} {2013})}\BibitemShut {NoStop}%
\bibitem [{\citenamefont {Keber}\ \emph {et~al.}(2014)\citenamefont {Keber},
  \citenamefont {Loiseau}, \citenamefont {Sanchez}, \citenamefont {DeCamp},
  \citenamefont {Giomi}, \citenamefont {Bowick}, \citenamefont {Marchetti},
  \citenamefont {Dogic},\ and\ \citenamefont {Bausch}}]{Keber:2014fh}%
  \BibitemOpen
  \bibfield  {author} {\bibinfo {author} {\bibfnamefont {F.~C.}\ \bibnamefont
  {Keber}}, \bibinfo {author} {\bibfnamefont {E.}~\bibnamefont {Loiseau}},
  \bibinfo {author} {\bibfnamefont {T.}~\bibnamefont {Sanchez}}, \bibinfo
  {author} {\bibfnamefont {S.~J.}\ \bibnamefont {DeCamp}}, \bibinfo {author}
  {\bibfnamefont {L.}~\bibnamefont {Giomi}}, \bibinfo {author} {\bibfnamefont
  {M.~J.}\ \bibnamefont {Bowick}}, \bibinfo {author} {\bibfnamefont {M.~C.}\
  \bibnamefont {Marchetti}}, \bibinfo {author} {\bibfnamefont {Z.}~\bibnamefont
  {Dogic}}, \ and\ \bibinfo {author} {\bibfnamefont {A.~R.}\ \bibnamefont
  {Bausch}},\ }\href@noop {} {\bibfield  {journal} {\bibinfo  {journal}
  {Science}\ }\textbf {\bibinfo {volume} {345}},\ \bibinfo
  {pages} {1135} (\bibinfo {year} {2014})}\BibitemShut {NoStop}%
\bibitem [{\citenamefont {Pearce}\ \emph {et~al.}(2018)\citenamefont {Pearce},
  \citenamefont {Ellis}, \citenamefont {Fernandez-Nieves},\ and\ \citenamefont
  {Giomi}}]{Pearce:2018wt}%
  \BibitemOpen
  \bibfield  {author} {\bibinfo {author} {\bibfnamefont {D.~J.~G.}\
  \bibnamefont {Pearce}}, \bibinfo {author} {\bibfnamefont {P.~W.}\
  \bibnamefont {Ellis}}, \bibinfo {author} {\bibfnamefont {A.}~\bibnamefont
  {Fernandez-Nieves}}, \ and\ \bibinfo {author} {\bibfnamefont
  {L.}~\bibnamefont {Giomi}},\ }\href@noop {} {\bibfield  {journal} {\bibinfo
  {journal} {arXiv}\ } (\bibinfo {year} {2018})},\ \Eprint
  {http://arxiv.org/abs/1805.01455v1} {1805.01455v1} \BibitemShut {NoStop}%
\bibitem [{\citenamefont {Voituriez}\ \emph {et~al.}(2007)\citenamefont
  {Voituriez}, \citenamefont {Joanny},\ and\ \citenamefont
  {Prost}}]{Voituriez:2007jy}%
  \BibitemOpen
  \bibfield  {author} {\bibinfo {author} {\bibfnamefont {R.}~\bibnamefont
  {Voituriez}}, \bibinfo {author} {\bibfnamefont {J.~F.}\ \bibnamefont
  {Joanny}}, \ and\ \bibinfo {author} {\bibfnamefont {J.}~\bibnamefont
  {Prost}},\ }\href@noop {} {\bibfield  {journal} {\bibinfo  {journal} {EPL
  (Europhysics Letters)}\ }\textbf {\bibinfo {volume} {70}},\ \bibinfo {pages}
  {404} (\bibinfo {year} {2007})}\BibitemShut {NoStop}%
\bibitem [{\citenamefont {F{\"u}rthauer}\ \emph {et~al.}(2012)\citenamefont
  {F{\"u}rthauer}, \citenamefont {Neef}, \citenamefont {Grill}, \citenamefont
  {Kruse},\ and\ \citenamefont {J{\"u}licher}}]{Furthauer:2012iu}%
  \BibitemOpen
  \bibfield  {author} {\bibinfo {author} {\bibfnamefont {S.}~\bibnamefont
  {F{\"u}rthauer}}, \bibinfo {author} {\bibfnamefont {M.}~\bibnamefont {Neef}},
  \bibinfo {author} {\bibfnamefont {S.~W.}\ \bibnamefont {Grill}}, \bibinfo
  {author} {\bibfnamefont {K.}~\bibnamefont {Kruse}}, \ and\ \bibinfo {author}
  {\bibfnamefont {F.}~\bibnamefont {J{\"u}licher}},\ }\href@noop {} {\bibfield
  {journal} {\bibinfo  {journal} {New J. Phys.}\ }\textbf {\bibinfo {volume}
  {14}},\ \bibinfo {pages} {023001} (\bibinfo {year} {2012})}\BibitemShut
  {NoStop}%
\bibitem [{\citenamefont {Duclos}\ \emph {et~al.}(2018)\citenamefont {Duclos},
  \citenamefont {Blanch-Mercader}, \citenamefont {Yashunsky}, \citenamefont
  {Salbreux}, \citenamefont {Joanny}, \citenamefont {Prost},\ and\
  \citenamefont {Silberzan}}]{Duclos:2018it}%
  \BibitemOpen
  \bibfield  {author} {\bibinfo {author} {\bibfnamefont {G.}~\bibnamefont
  {Duclos}}, \bibinfo {author} {\bibfnamefont {C.}~\bibnamefont
  {Blanch-Mercader}}, \bibinfo {author} {\bibfnamefont {V.}~\bibnamefont
  {Yashunsky}}, \bibinfo {author} {\bibfnamefont {G.}~\bibnamefont {Salbreux}},
  \bibinfo {author} {\bibfnamefont {J.~F.}\ \bibnamefont {Joanny}}, \bibinfo
  {author} {\bibfnamefont {J.}~\bibnamefont {Prost}}, \ and\ \bibinfo {author}
  {\bibfnamefont {P.}~\bibnamefont {Silberzan}},\ }\href@noop {} {\bibfield
  {journal} {\bibinfo  {journal} {Nat. Phys.}\ }\textbf {\bibinfo {volume}
  {14}},\ \bibinfo {pages} {728} (\bibinfo {year} {2018})}\BibitemShut
  {NoStop}%
\bibitem [{\citenamefont {Hannezo}\ \emph {et~al.}(2014)\citenamefont
  {Hannezo}, \citenamefont {Prost},\ and\ \citenamefont
  {Joanny}}]{Hannezo:2014gp}%
  \BibitemOpen
  \bibfield  {author} {\bibinfo {author} {\bibfnamefont {E.}~\bibnamefont
  {Hannezo}}, \bibinfo {author} {\bibfnamefont {J.}~\bibnamefont {Prost}}, \
  and\ \bibinfo {author} {\bibfnamefont {J.-F.}\ \bibnamefont {Joanny}},\
  }\href@noop {} {\bibfield  {journal} {\bibinfo  {journal} {Proc. Natl. Acad. Sci. USA}\ }\textbf
  {\bibinfo {volume} {111}},\ \bibinfo {pages} {27} (\bibinfo {year}
  {2014})}\BibitemShut {NoStop}%
\bibitem [{\citenamefont {Ideses}\ \emph {et~al.}(2018)\citenamefont {Ideses},
  \citenamefont {Erukhimovitch}, \citenamefont {Brand}, \citenamefont
  {Jourdain}, \citenamefont {Hernandez}, \citenamefont {Gabineti},
  \citenamefont {Safran}, \citenamefont {Kruse},\ and\ \citenamefont
  {Bernheim-Groswasser}}]{Ideses:2018dn}%
  \BibitemOpen
  \bibfield  {author} {\bibinfo {author} {\bibfnamefont {Y.}~\bibnamefont
  {Ideses}}, \bibinfo {author} {\bibfnamefont {V.}~\bibnamefont
  {Erukhimovitch}}, \bibinfo {author} {\bibfnamefont {R.}~\bibnamefont
  {Brand}}, \bibinfo {author} {\bibfnamefont {D.}~\bibnamefont {Jourdain}},
  \bibinfo {author} {\bibfnamefont {J.~S.}\ \bibnamefont {Hernandez}}, \bibinfo
  {author} {\bibfnamefont {U.~R.}\ \bibnamefont {Gabineti}}, \bibinfo {author}
  {\bibfnamefont {S.~A.}\ \bibnamefont {Safran}}, \bibinfo {author}
  {\bibfnamefont {K.}~\bibnamefont {Kruse}}, \ and\ \bibinfo {author}
  {\bibfnamefont {A.}~\bibnamefont {Bernheim-Groswasser}},\ }\href@noop {}
  {\bibfield  {journal} {\bibinfo  {journal} {Nat. Comms.}\ }\textbf
  {\bibinfo {volume} {9}},\ \bibinfo {pages} {2461} (\bibinfo {year} {2018})}\BibitemShut {NoStop}%
\bibitem [{\citenamefont {Fritzsche}\ \emph {et~al.}(2016)\citenamefont
  {Fritzsche}, \citenamefont {Erlenk{\"a}mper}, \citenamefont {Moeendarbary},
  \citenamefont {Charras},\ and\ \citenamefont {Kruse}}]{Fritzsche:2016fx}%
  \BibitemOpen
  \bibfield  {author} {\bibinfo {author} {\bibfnamefont {M.}~\bibnamefont
  {Fritzsche}}, \bibinfo {author} {\bibfnamefont {C.}~\bibnamefont
  {Erlenk{\"a}mper}}, \bibinfo {author} {\bibfnamefont {E.}~\bibnamefont
  {Moeendarbary}}, \bibinfo {author} {\bibfnamefont {G.}~\bibnamefont
  {Charras}}, \ and\ \bibinfo {author} {\bibfnamefont {K.}~\bibnamefont
  {Kruse}},\ }\href@noop {} {\bibfield  {journal} {\bibinfo  {journal} {Sci.
  Adv.}\ }\textbf {\bibinfo {volume} {2}},\ \bibinfo {pages} {e1501337}
  (\bibinfo {year} {2016})}\BibitemShut {NoStop}%
\bibitem [{\citenamefont {Chugh}\ \emph {et~al.}(2017)\citenamefont {Chugh},
  \citenamefont {Clark}, \citenamefont {Smith}, \citenamefont {Cassani},
  \citenamefont {Dierkes}, \citenamefont {Ragab}, \citenamefont {Roux},
  \citenamefont {Charras}, \citenamefont {Salbreux},\ and\ \citenamefont
  {Paluch}}]{Chugh:2017de}%
  \BibitemOpen
  \bibfield  {author} {\bibinfo {author} {\bibfnamefont {P.}~\bibnamefont
  {Chugh}}, \bibinfo {author} {\bibfnamefont {A.~G.}\ \bibnamefont {Clark}},
  \bibinfo {author} {\bibfnamefont {M.~B.}\ \bibnamefont {Smith}}, \bibinfo
  {author} {\bibfnamefont {D.~A.~D.}\ \bibnamefont {Cassani}}, \bibinfo
  {author} {\bibfnamefont {K.}~\bibnamefont {Dierkes}}, \bibinfo {author}
  {\bibfnamefont {A.}~\bibnamefont {Ragab}}, \bibinfo {author} {\bibfnamefont
  {P.~P.}\ \bibnamefont {Roux}}, \bibinfo {author} {\bibfnamefont
  {G.}~\bibnamefont {Charras}}, \bibinfo {author} {\bibfnamefont
  {G.}~\bibnamefont {Salbreux}}, \ and\ \bibinfo {author} {\bibfnamefont
  {E.~K.}\ \bibnamefont {Paluch}},\ }\href@noop {} {\bibfield  {journal}
  {\bibinfo  {journal} {Nat. Cell Biol.}\ }\textbf {\bibinfo {volume} {19}},\
  \bibinfo {pages} {689} (\bibinfo {year} {2017})}\BibitemShut {NoStop}%
\bibitem [{\citenamefont {Salbreux}\ \emph {et~al.}(2012)\citenamefont
  {Salbreux}, \citenamefont {Charras},\ and\ \citenamefont
  {Paluch}}]{Salbreux:2012bo}%
  \BibitemOpen
  \bibfield  {author} {\bibinfo {author} {\bibfnamefont {G.}~\bibnamefont
  {Salbreux}}, \bibinfo {author} {\bibfnamefont {G.}~\bibnamefont {Charras}}, \
  and\ \bibinfo {author} {\bibfnamefont {E.}~\bibnamefont {Paluch}},\
  }\href@noop {} {\bibfield  {journal} {\bibinfo  {journal} {Trends Cell
  Biol.}\ }\textbf {\bibinfo {volume} {22}},\ \bibinfo {pages} {536}
  (\bibinfo {year} {2012})}\BibitemShut {NoStop}%
\bibitem [{\citenamefont {Blanchoin}\ \emph {et~al.}(2014)\citenamefont
  {Blanchoin}, \citenamefont {Boujemaa-Paterski}, \citenamefont {Sykes},\ and\
  \citenamefont {Plastino}}]{Blanchoin:2014jr}%
  \BibitemOpen
  \bibfield  {author} {\bibinfo {author} {\bibfnamefont {L.}~\bibnamefont
  {Blanchoin}}, \bibinfo {author} {\bibfnamefont {R.}~\bibnamefont
  {Boujemaa-Paterski}}, \bibinfo {author} {\bibfnamefont {C.}~\bibnamefont
  {Sykes}}, \ and\ \bibinfo {author} {\bibfnamefont {J.}~\bibnamefont
  {Plastino}},\ }\href@noop {} {\bibfield  {journal} {\bibinfo  {journal}
  {Physiol. Rev.}\ }\textbf {\bibinfo {volume} {94}},\ \bibinfo {pages}
  {235} (\bibinfo {year} {2014})}\BibitemShut {NoStop}%
\bibitem [{\citenamefont {Zumdieck}\ \emph {et~al.}(2005)\citenamefont
  {Zumdieck}, \citenamefont {Cosentino~Lagomarsino}, \citenamefont {Tanase},
  \citenamefont {Kruse}, \citenamefont {Mulder}, \citenamefont {Dogterom},\
  and\ \citenamefont {J{\"u}licher}}]{Zumdieck:2005wba}%
  \BibitemOpen
  \bibfield  {author} {\bibinfo {author} {\bibfnamefont {A.}~\bibnamefont
  {Zumdieck}}, \bibinfo {author} {\bibfnamefont {M.}~\bibnamefont
  {Cosentino~Lagomarsino}}, \bibinfo {author} {\bibfnamefont {C.}~\bibnamefont
  {Tanase}}, \bibinfo {author} {\bibfnamefont {K.}~\bibnamefont {Kruse}},
  \bibinfo {author} {\bibfnamefont {B.}~\bibnamefont {Mulder}}, \bibinfo
  {author} {\bibfnamefont {M.}~\bibnamefont {Dogterom}}, \ and\ \bibinfo
  {author} {\bibfnamefont {F.}~\bibnamefont {J{\"u}licher}},\ }\href@noop {}
  {\bibfield  {journal} {\bibinfo  {journal} {Phys. Rev. Lett.}\
  }\textbf {\bibinfo {volume} {95}},\ \bibinfo {pages} {258103} (\bibinfo
  {year} {2005})}\BibitemShut {NoStop}%
\bibitem [{\citenamefont {Salbreux}\ \emph {et~al.}(2009)\citenamefont
  {Salbreux}, \citenamefont {Prost},\ and\ \citenamefont
  {Joanny}}]{Salbreux:2009fp}%
  \BibitemOpen
  \bibfield  {author} {\bibinfo {author} {\bibfnamefont {G.}~\bibnamefont
  {Salbreux}}, \bibinfo {author} {\bibfnamefont {J.}~\bibnamefont {Prost}}, \
  and\ \bibinfo {author} {\bibfnamefont {J.~F.}\ \bibnamefont {Joanny}},\
  }\href@noop {} {\bibfield  {journal} {\bibinfo  {journal} {Phys. Rev. Lett.}\
  }\textbf {\bibinfo {volume} {103}},\ \bibinfo {pages} {058102} (\bibinfo
  {year} {2009})}\BibitemShut {NoStop}%
\bibitem [{\citenamefont {Sedzinski}\ \emph {et~al.}(2011)\citenamefont
  {Sedzinski}, \citenamefont {Biro}, \citenamefont {Oswald}, \citenamefont
  {Tinevez}, \citenamefont {Salbreux},\ and\ \citenamefont
  {Paluch}}]{Sedzinski:2011ef}%
  \BibitemOpen
  \bibfield  {author} {\bibinfo {author} {\bibfnamefont {J.}~\bibnamefont
  {Sedzinski}}, \bibinfo {author} {\bibfnamefont {M.}~\bibnamefont {Biro}},
  \bibinfo {author} {\bibfnamefont {A.}~\bibnamefont {Oswald}}, \bibinfo
  {author} {\bibfnamefont {J.-Y.}\ \bibnamefont {Tinevez}}, \bibinfo {author}
  {\bibfnamefont {G.}~\bibnamefont {Salbreux}}, \ and\ \bibinfo {author}
  {\bibfnamefont {E.}~\bibnamefont {Paluch}},\ }\href@noop {} {\bibfield
  {journal} {\bibinfo  {journal} {Nature}\ }\textbf {\bibinfo {volume} {476}},\
  \bibinfo {pages} {462} (\bibinfo {year} {2011})}\BibitemShut {NoStop}%
\bibitem [{\citenamefont {Turlier}\ \emph {et~al.}(2014)\citenamefont
  {Turlier}, \citenamefont {Audoly}, \citenamefont {Prost},\ and\ \citenamefont
  {Joanny}}]{Turlier:2014hq}%
  \BibitemOpen
  \bibfield  {author} {\bibinfo {author} {\bibfnamefont {H.}~\bibnamefont
  {Turlier}}, \bibinfo {author} {\bibfnamefont {B.}~\bibnamefont {Audoly}},
  \bibinfo {author} {\bibfnamefont {J.}~\bibnamefont {Prost}}, \ and\ \bibinfo
  {author} {\bibfnamefont {J.-F.}\ \bibnamefont {Joanny}},\ }\href@noop {}
  {\bibfield  {journal} {\bibinfo  {journal} {Biophys. J.}\ }\textbf {\bibinfo
  {volume} {106}},\ \bibinfo {pages} {114} (\bibinfo {year}
  {2014})}\BibitemShut {NoStop}%
\bibitem [{\citenamefont {Clark}\ \emph {et~al.}(2013)\citenamefont {Clark},
  \citenamefont {Dierkes},\ and\ \citenamefont {Paluch}}]{Clark:2013ef}%
  \BibitemOpen
  \bibfield  {author} {\bibinfo {author} {\bibfnamefont {A.~G.}\ \bibnamefont
  {Clark}}, \bibinfo {author} {\bibfnamefont {K.}~\bibnamefont {Dierkes}}, \
  and\ \bibinfo {author} {\bibfnamefont {E.~K.}\ \bibnamefont {Paluch}},\
  }\href@noop {} {\bibfield  {journal} {\bibinfo  {journal} {Biophys J}\
  }\textbf {\bibinfo {volume} {105}},\ \bibinfo {pages} {570} (\bibinfo {year}
  {2013})}\BibitemShut {NoStop}%
\bibitem [{\citenamefont {Clausen}\ \emph {et~al.}(2017)\citenamefont
  {Clausen}, \citenamefont {Colin-York}, \citenamefont {Schneider},
  \citenamefont {Eggeling},\ and\ \citenamefont {Fritzsche}}]{Clausen:2017jc}%
  \BibitemOpen
  \bibfield  {author} {\bibinfo {author} {\bibfnamefont {M.~P.}\ \bibnamefont
  {Clausen}}, \bibinfo {author} {\bibfnamefont {H.}~\bibnamefont {Colin-York}},
  \bibinfo {author} {\bibfnamefont {F.}~\bibnamefont {Schneider}}, \bibinfo
  {author} {\bibfnamefont {C.}~\bibnamefont {Eggeling}}, \ and\ \bibinfo
  {author} {\bibfnamefont {M.}~\bibnamefont {Fritzsche}},\ }\href@noop {}
  {\bibfield  {journal} {\bibinfo  {journal} {J. Phys. D}\
  }\textbf {\bibinfo {volume} {50}},\ \bibinfo {pages} {064002} (\bibinfo
  {year} {2017})}\BibitemShut {NoStop}%
\bibitem [{\citenamefont {Joanny}\ \emph {et~al.}(2013)\citenamefont {Joanny},
  \citenamefont {Kruse}, \citenamefont {Prost},\ and\ \citenamefont
  {Ramaswamy}}]{Joanny:2013fm}%
  \BibitemOpen
  \bibfield  {author} {\bibinfo {author} {\bibfnamefont {J.~F.}\ \bibnamefont
  {Joanny}}, \bibinfo {author} {\bibfnamefont {K.}~\bibnamefont {Kruse}},
  \bibinfo {author} {\bibfnamefont {J.}~\bibnamefont {Prost}}, \ and\ \bibinfo
  {author} {\bibfnamefont {S.}~\bibnamefont {Ramaswamy}},\ }\href@noop {}
  {\bibfield  {journal} {\bibinfo  {journal} {Eur. Phys. J. E}\ }\textbf {\bibinfo
  {volume} {36}},\ \bibinfo {pages} {52} (\bibinfo {year} {2013})}\BibitemShut
  {NoStop}%
\bibitem [{\citenamefont {Mayer}\ \emph {et~al.}(2010)\citenamefont {Mayer},
  \citenamefont {Depken}, \citenamefont {Bois}, \citenamefont {J{\"u}licher},\
  and\ \citenamefont {Grill}}]{Mayer:2010kt}%
  \BibitemOpen
  \bibfield  {author} {\bibinfo {author} {\bibfnamefont {M.}~\bibnamefont
  {Mayer}}, \bibinfo {author} {\bibfnamefont {M.}~\bibnamefont {Depken}},
  \bibinfo {author} {\bibfnamefont {J.~S.}\ \bibnamefont {Bois}}, \bibinfo
  {author} {\bibfnamefont {F.}~\bibnamefont {J{\"u}licher}}, \ and\ \bibinfo
  {author} {\bibfnamefont {S.~W.}\ \bibnamefont {Grill}},\ }\href@noop {}
  {\bibfield  {journal} {\bibinfo  {journal} {Nature}\ }\textbf {\bibinfo
  {volume} {467}},\ \bibinfo {pages} {617} (\bibinfo {year}
  {2010})}\BibitemShut {NoStop}%
\bibitem [{\citenamefont {Naganathan}\ \emph {et~al.}(2014)\citenamefont
  {Naganathan}, \citenamefont {F{\"u}rthauer}, \citenamefont {Nishikawa},
  \citenamefont {J{\"u}licher},\ and\ \citenamefont
  {Grill}}]{Naganathan:2014fc}%
  \BibitemOpen
  \bibfield  {author} {\bibinfo {author} {\bibfnamefont {S.~R.}\ \bibnamefont
  {Naganathan}}, \bibinfo {author} {\bibfnamefont {S.}~\bibnamefont
  {F{\"u}rthauer}}, \bibinfo {author} {\bibfnamefont {M.}~\bibnamefont
  {Nishikawa}}, \bibinfo {author} {\bibfnamefont {F.}~\bibnamefont
  {J{\"u}licher}}, \ and\ \bibinfo {author} {\bibfnamefont {S.~W.}\
  \bibnamefont {Grill}},\ }\href@noop {} {\bibfield  {journal} {\bibinfo
  {journal} {eLife}\ }\textbf {\bibinfo {volume} {3}},\ \bibinfo {pages}
  {e04165} (\bibinfo {year} {2014})}\BibitemShut {NoStop}%
\bibitem [{\citenamefont {Berthoumieux}\ \emph {et~al.}(2014)\citenamefont
  {Berthoumieux}, \citenamefont {Maitre}, \citenamefont {Heisenberg},
  \citenamefont {Paluch}, \citenamefont {J\"ulicher},\ and\ \citenamefont
  {Salbreux}}]{Berthoumieux:2014eo}%
  \BibitemOpen
  \bibfield  {author} {\bibinfo {author} {\bibfnamefont {H.}~\bibnamefont
  {Berthoumieux}}, \bibinfo {author} {\bibfnamefont {J.-L.}\ \bibnamefont
  {Maitre}}, \bibinfo {author} {\bibfnamefont {C.-P.}\ \bibnamefont
  {Heisenberg}}, \bibinfo {author} {\bibfnamefont {E.~K.}\ \bibnamefont
  {Paluch}}, \bibinfo {author} {\bibfnamefont {F.}~\bibnamefont {J\"ulicher}}, \
  and\ \bibinfo {author} {\bibfnamefont {G.}~\bibnamefont {Salbreux}},\
  }\href@noop {} {\bibfield  {journal} {\bibinfo  {journal} {New J. Phys.}\
  }\textbf {\bibinfo {volume} {16}},\ \bibinfo {pages} {065005} (\bibinfo {year} {2014})}\BibitemShut
  {NoStop}%
\bibitem [{\citenamefont {Mietke}\ \emph {et~al.}(2019)\citenamefont {Mietke},
  \citenamefont {J{\"u}licher},\ and\ \citenamefont
  {Sbalzarini}}]{Mietke:2019ki}%
  \BibitemOpen
  \bibfield  {author} {\bibinfo {author} {\bibfnamefont {A.}~\bibnamefont
  {Mietke}}, \bibinfo {author} {\bibfnamefont {F.}~\bibnamefont
  {J{\"u}licher}}, \ and\ \bibinfo {author} {\bibfnamefont {I.~F.}\
  \bibnamefont {Sbalzarini}},\ }\href@noop {} {\bibfield  {journal} {\bibinfo
  {journal} {Proc. Natl. Acad. Sci. USA}\ }\textbf {\bibinfo {volume} {116}},\ \bibinfo {pages} {29}
  (\bibinfo {year} {2019})}\BibitemShut {NoStop}%
\bibitem [{\citenamefont {Bois}\ \emph {et~al.}(2011)\citenamefont {Bois},
  \citenamefont {J{\"u}licher},\ and\ \citenamefont {Grill}}]{Bois:2011kx}%
  \BibitemOpen
  \bibfield  {author} {\bibinfo {author} {\bibfnamefont {J.~S.}\ \bibnamefont
  {Bois}}, \bibinfo {author} {\bibfnamefont {F.}~\bibnamefont {J{\"u}licher}},
  \ and\ \bibinfo {author} {\bibfnamefont {S.~W.}\ \bibnamefont {Grill}},\
  }\href@noop {} {\bibfield  {journal} {\bibinfo  {journal} {Phys. Rev.
  Lett.}\ }\textbf {\bibinfo {volume} {106}},\ \bibinfo {pages} {028103}
  (\bibinfo {year} {2011})}\BibitemShut {NoStop}%
\bibitem [{\citenamefont {Hannezo}\ \emph {et~al.}(2015)\citenamefont
  {Hannezo}, \citenamefont {Dong}, \citenamefont {Recho}, \citenamefont
  {Joanny},\ and\ \citenamefont {Hayashi}}]{Hannezo:2015ba}%
  \BibitemOpen
  \bibfield  {author} {\bibinfo {author} {\bibfnamefont {E.}~\bibnamefont
  {Hannezo}}, \bibinfo {author} {\bibfnamefont {B.}~\bibnamefont {Dong}},
  \bibinfo {author} {\bibfnamefont {P.}~\bibnamefont {Recho}}, \bibinfo
  {author} {\bibfnamefont {J.-F.}\ \bibnamefont {Joanny}}, \ and\ \bibinfo
  {author} {\bibfnamefont {S.}~\bibnamefont {Hayashi}},\ }\href@noop {}
  {\bibfield  {journal} {\bibinfo  {journal} {Proc.~Natl.~Acad.~Sci.~USA}\ }\textbf {\bibinfo
  {volume} {112}},\ \bibinfo {pages} {8620} (\bibinfo {year}
  {2015})}\BibitemShut {NoStop}%
\bibitem{SM}
See Supplemental Material at [URL will be inserted by publisher] for movies of the numerical solutions.
\bibitem [{\citenamefont {Aranson}\ and\ \citenamefont
  {Kramer}(2002)}]{Aranson:2002vf}%
  \BibitemOpen
  \bibfield  {author} {\bibinfo {author} {\bibfnamefont {I.~S.}\ \bibnamefont
  {Aranson}}\ and\ \bibinfo {author} {\bibfnamefont {L.}~\bibnamefont
  {Kramer}},\ }\href@noop {} {\bibfield  {journal} {\bibinfo  {journal} {Rev. Mod. Phys.}\ }\textbf {\bibinfo {volume} {74}},\ \bibinfo
  {pages} {99} (\bibinfo {year} {2002})}\BibitemShut {NoStop}%
\bibitem [{\citenamefont {Tribelsky}\ and\ \citenamefont
  {Tsuboi}(1996)}]{Tribelsky:1996do}%
  \BibitemOpen
  \bibfield  {author} {\bibinfo {author} {\bibfnamefont {M.~I.}~\bibnamefont
  {Tribelsky}}\ and\ \bibinfo {author} {\bibfnamefont {K.}~\bibnamefont
  {Tsuboi}},\ }\href@noop {} {\bibfield  {journal} {\bibinfo  {journal}
  {Phys. Rev. Lett.}\ }\textbf {\bibinfo {volume} {76}},\ \bibinfo
  {pages} {1631} (\bibinfo {year} {1996})}\BibitemShut {NoStop}%
\bibitem [{\citenamefont {Kruse}\ and\ \citenamefont
  {J{\"u}licher}(2003)}]{Kruse:2003vr}%
  \BibitemOpen
  \bibfield  {author} {\bibinfo {author} {\bibfnamefont {K.}~\bibnamefont
  {Kruse}}\ and\ \bibinfo {author} {\bibfnamefont {F.}~\bibnamefont
  {J{\"u}licher}},\ }\href@noop {} {\bibfield  {journal} {\bibinfo  {journal}
  {Phys. Rev. E}\ }\textbf {\bibinfo {volume}
  {67}},\ \bibinfo {pages} {051913} (\bibinfo {year} {2003})}\BibitemShut
  {NoStop}%
\bibitem [{\citenamefont {Gowrishankar}\ \emph {et~al.}(2012)\citenamefont
  {Gowrishankar}, \citenamefont {Ghosh}, \citenamefont {Saha}, \citenamefont
  {C}, \citenamefont {Mayor},\ and\ \citenamefont {Rao}}]{Gowrishankar:2012ek}%
  \BibitemOpen
  \bibfield  {author} {\bibinfo {author} {\bibfnamefont {K.}~\bibnamefont
  {Gowrishankar}}, \bibinfo {author} {\bibfnamefont {S.}~\bibnamefont {Ghosh}},
  \bibinfo {author} {\bibfnamefont {S.}~\bibnamefont {Saha}}, \bibinfo {author}
  {\bibfnamefont {R.}~\bibnamefont {C}}, \bibinfo {author} {\bibfnamefont
  {S.}~\bibnamefont {Mayor}}, \ and\ \bibinfo {author} {\bibfnamefont
  {M.}~\bibnamefont {Rao}},\ }\href@noop {} {\bibfield  {journal} {\bibinfo
  {journal} {Cell}\ }\textbf {\bibinfo {volume} {149}},\ \bibinfo {pages}
  {1353} (\bibinfo {year} {2012})}\BibitemShut {NoStop}%
\bibitem [{\citenamefont {Kumar}\ \emph {et~al.}(2014)\citenamefont {Kumar},
  \citenamefont {Bois}, \citenamefont {J\"ulicher},\ and\ \citenamefont
  {Grill}}]{Kumar:2014fx}%
  \BibitemOpen
  \bibfield  {author} {\bibinfo {author} {\bibfnamefont {K.~V.}\ \bibnamefont
  {Kumar}}, \bibinfo {author} {\bibfnamefont {J.~S.}\ \bibnamefont {Bois}},
  \bibinfo {author} {\bibfnamefont {F.}~\bibnamefont {J\"ulicher}}, \ and\
  \bibinfo {author} {\bibfnamefont {S.~W.}\ \bibnamefont {Grill}},\ }\href@noop
  {} {\bibfield  {journal} {\bibinfo  {journal} {Phys. Rev. Lett.}\ }\textbf
  {\bibinfo {volume} {112}},\ \bibinfo {pages} {208101} (\bibinfo {year} {2014})}\BibitemShut {NoStop}%
\bibitem [{\citenamefont {Banerjee}\ \emph {et~al.}(2017)\citenamefont
  {Banerjee}, \citenamefont {Munjal}, \citenamefont {Lecuit},\ and\
  \citenamefont {Rao}}]{Banerjee:2017ja}%
  \BibitemOpen
  \bibfield  {author} {\bibinfo {author} {\bibfnamefont {D.~S.}\ \bibnamefont
  {Banerjee}}, \bibinfo {author} {\bibfnamefont {A.}~\bibnamefont {Munjal}},
  \bibinfo {author} {\bibfnamefont {T.}~\bibnamefont {Lecuit}}, \ and\ \bibinfo
  {author} {\bibfnamefont {M.}~\bibnamefont {Rao}},\ }\href@noop {} {\bibfield
  {journal} {\bibinfo  {journal} {Nat. Comms.}\ }\textbf {\bibinfo
  {volume} {8}},\ \bibinfo {pages} {1121} (\bibinfo {year} {2017})}\BibitemShut
  {NoStop}%
\bibitem [{\citenamefont {Dombrowski}\ \emph {et~al.}(2004)\citenamefont
  {Dombrowski}, \citenamefont {Cisneros}, \citenamefont {Chatkaew},
  \citenamefont {Goldstein},\ and\ \citenamefont
  {Kessler}}]{Dombrowski:2004eu}%
  \BibitemOpen
  \bibfield  {author} {\bibinfo {author} {\bibfnamefont {C.}~\bibnamefont
  {Dombrowski}}, \bibinfo {author} {\bibfnamefont {L.}~\bibnamefont
  {Cisneros}}, \bibinfo {author} {\bibfnamefont {S.}~\bibnamefont {Chatkaew}},
  \bibinfo {author} {\bibfnamefont {R.~E.}\ \bibnamefont {Goldstein}}, \ and\
  \bibinfo {author} {\bibfnamefont {J.~O.}\ \bibnamefont {Kessler}},\
  }\href@noop {} {\bibfield  {journal} {\bibinfo  {journal} {Phys. Rev. Lett.}\
  }\textbf {\bibinfo {volume} {93}},\ \bibinfo {pages} {098103} (\bibinfo {year} {2004})}\BibitemShut
  {NoStop}%
\bibitem [{\citenamefont {Neef}\ and\ \citenamefont
  {Kruse}(2014)}]{Neef:2014ez}%
  \BibitemOpen
  \bibfield  {author} {\bibinfo {author} {\bibfnamefont {M.}~\bibnamefont
  {Neef}}\ and\ \bibinfo {author} {\bibfnamefont {K.}~\bibnamefont {Kruse}},\
  }\href@noop {} {\bibfield  {journal} {\bibinfo  {journal} {Phys. Rev. E}\ }\textbf {\bibinfo {volume} {90}},\ \bibinfo {pages} {052703}
   (\bibinfo {year}
  {2014})}\BibitemShut {NoStop}%
\bibitem [{\citenamefont {Ramaswamy}\ and\ \citenamefont
  {J{\"u}licher}(2016)}]{Ramaswamy:2016kb}%
  \BibitemOpen
  \bibfield  {author} {\bibinfo {author} {\bibfnamefont {R.}~\bibnamefont
  {Ramaswamy}}\ and\ \bibinfo {author} {\bibfnamefont {F.}~\bibnamefont
  {J{\"u}licher}},\ }\href@noop {} {\bibfield  {journal} {\bibinfo  {journal}
  {Sci. Rep.}\ }\textbf {\bibinfo {volume} {6}},\ \bibinfo {pages} {20838}
  (\bibinfo {year} {2016})}\BibitemShut {NoStop}%
\bibitem [{\citenamefont {Doostmohammadi}\ \emph {et~al.}(2016)\citenamefont
  {Doostmohammadi}, \citenamefont {Adamer}, \citenamefont {Thampi},\ and\
  \citenamefont {Yeomans}}]{Doostmohammadi:2016bd}%
  \BibitemOpen
  \bibfield  {author} {\bibinfo {author} {\bibfnamefont {A.}~\bibnamefont
  {Doostmohammadi}}, \bibinfo {author} {\bibfnamefont {M.~F.}\ \bibnamefont
  {Adamer}}, \bibinfo {author} {\bibfnamefont {S.~P.}\ \bibnamefont {Thampi}},
  \ and\ \bibinfo {author} {\bibfnamefont {J.~M.}\ \bibnamefont {Yeomans}},\
  }\href@noop {} {\bibfield  {journal} {\bibinfo  {journal} {Nat.
  Comms.}\ }\textbf {\bibinfo {volume} {7}},\ \bibinfo {pages} {10557}
  (\bibinfo {year} {2016})}\BibitemShut {NoStop}%
  \bibitem [{\citenamefont {Fritzsche}\ \emph {et~al.}(2017)\citenamefont
  {Fritzsche}, \citenamefont {Fernandes}, \citenamefont {Chang}, \citenamefont
  {Colin-York}, \citenamefont {Clausen}, \citenamefont {Felce}, \citenamefont
  {Galiani}, \citenamefont {Erlenk{\"a}mper}, \citenamefont {Santos},
  \citenamefont {Heddleston}, \citenamefont {Pedroza-Pacheco}, \citenamefont
  {Waithe}, \citenamefont {de~la Serna}, \citenamefont {Lagerholm},
  \citenamefont {Liu}, \citenamefont {Chew}, \citenamefont {Betzig},
  \citenamefont {Davis},\ and\ \citenamefont {Eggeling}}]{Fritzsche:2017ji}%
  \BibitemOpen
  \bibfield  {author} {\bibinfo {author} {\bibfnamefont {M.}~\bibnamefont
  {Fritzsche}}, \bibinfo {author} {\bibfnamefont {R.~A.}\ \bibnamefont
  {Fernandes}}, \bibinfo {author} {\bibfnamefont {V.~T.}\ \bibnamefont
  {Chang}}, \bibinfo {author} {\bibfnamefont {H.}~\bibnamefont {Colin-York}},
  \bibinfo {author} {\bibfnamefont {M.~P.}\ \bibnamefont {Clausen}}, \bibinfo
  {author} {\bibfnamefont {J.~H.}\ \bibnamefont {Felce}}, \bibinfo {author}
  {\bibfnamefont {S.}~\bibnamefont {Galiani}}, \bibinfo {author} {\bibfnamefont
  {C.}~\bibnamefont {Erlenk{\"a}mper}}, \bibinfo {author} {\bibfnamefont
  {A.~M.}\ \bibnamefont {Santos}}, \bibinfo {author} {\bibfnamefont {J.~M.}\
  \bibnamefont {Heddleston}}, \bibinfo {author} {\bibfnamefont
  {I.}~\bibnamefont {Pedroza-Pacheco}}, \bibinfo {author} {\bibfnamefont
  {D.}~\bibnamefont {Waithe}}, \bibinfo {author} {\bibfnamefont {J.~B.}\
  \bibnamefont {de~la Serna}}, \bibinfo {author} {\bibfnamefont {B.~C.}\
  \bibnamefont {Lagerholm}}, \bibinfo {author} {\bibfnamefont {T.-L.}\
  \bibnamefont {Liu}}, \bibinfo {author} {\bibfnamefont {T.-L.}\ \bibnamefont
  {Chew}}, \bibinfo {author} {\bibfnamefont {E.}~\bibnamefont {Betzig}},
  \bibinfo {author} {\bibfnamefont {S.~J.}\ \bibnamefont {Davis}}, \ and\
  \bibinfo {author} {\bibfnamefont {C.}~\bibnamefont {Eggeling}},\ }\href@noop
  {} {\bibfield  {journal} {\bibinfo  {journal} {Sci Adv}\ }\textbf {\bibinfo
  {volume} {3}},\ \bibinfo {pages} {e1603032} (\bibinfo {year} {2017})}\BibitemShut {NoStop}%
\bibitem [{\citenamefont {Charras}(2008)}]{Charras:2008uu}%
  \BibitemOpen
  \bibfield  {author} {\bibinfo {author} {\bibfnamefont {G.~T.}\ \bibnamefont
  {Charras}},\ }\href@noop {} {\bibfield  {journal} {\bibinfo  {journal} {J
  Microsc-Oxford}\ }\textbf {\bibinfo {volume} {231}},\ \bibinfo {pages} {466}
  (\bibinfo {year} {2008})}\BibitemShut {NoStop}%
\end{thebibliography}
%

\end{document}